\documentclass[varenna,seceqno,cite]{cimento}
\usepackage[dvips]{graphicx}
\usepackage[english]{babel}
\usepackage{amsmath}
\usepackage{amssymb}
\usepackage{times,bm}
\newcommand{\bs}{\begin{split}}
\newcommand{\es}{\end{split}}
\newcommand{\bea}{\begin{eqnarray}}
\newcommand{\eea}{\end{eqnarray}}
\newcommand{\be}{\begin{equation}}
\newcommand{\ee}{\end{equation}}
\newcommand{\ba}{\begin{eqnarray}}
\newcommand{\ea}{\end{eqnarray}}
\newcommand{\ek}{\epsilon_{\mathbf{k}}}
\newcommand{\Ek}{E_{\mathbf{k}}}
\newcommand{\uk}{u_{\mathbf{k}}}
\newcommand{\vk}{v_{\mathbf{k}}}

\newcommand{\xik}{\xi_{\mathbf{k}}}

\newcommand{\phik}{\varphi_{\mathbf{k}}}

\begin{document}

\title{Finite Temperature Effects in Ultracold Fermi Gases}

\author{K. Levin \atque Qijin Chen}

\institute{James Franck Institute and Department of Physics,
 University of Chicago, Chicago, Illinois 60637, USA}



\maketitle





\section{BCS-BEC Crossover Theory and the Physical Effects of
Temperature}
\label{sec:I}

The study of ultracold trapped fermionic gases is a rapidly exploding
subject \cite{Jin3,Grimm,Jin4,Ketterle3,KetterleV,
Thomas2,Grimm3,ThermoScience,Salomon3,Hulet4} which is defining new directions
in condensed matter and atomic physics.  It has also captured the
attention of physicists who study color superconducting quark matter as
well as nuclear matter \cite{Wilczek,LW03,FGLW05}.  Indeed, it is hard,
in recent times, to find a subfield of physics which appeals this
broadly to the research community.  As we come to understand the gases,
experimentalists will move to address fermionic atoms in optical
lattices; this will provide important insight to condensed matter
physicists as analogue systems for ``solving" intractable many body
problems.

What makes these gases (and lattices) so important is their remarkable
tunability and controllability. Using a Feshbach resonance, one can tune
the attractive two-body interaction from weak to strong, and thereby
make a smooth crossover from a BCS superfluid to a Bose-Einstein
condensation (BEC) \cite{Leggett,Eagles}.  Not only does this allow high
transition temperatures $T_c$ (relative to the Fermi energy $E_F$) but
it may also provide insights into the high temperature cuprate
superconductivity
\cite{Chen2,ourreview,ReviewJLTP,LeggettNature}. Furthermore, one can
tune the population of the two spin states, essentially at will, in this
way, allowing exploration \cite{ZSSK06,PLKLH06,ZSSK206} of exotic
polarized phases such as the FFLO \cite{FFLO,LOFF_Review,Combescot}
superfluids, of interest to condensed matter, nuclear and particle
physicists. One will be able to tune the lattice parameters such as
bandwidth, on-site repulsion, even random disorder, etc and thereby
study the famous, and as yet, unsolved Hubbard model Hamiltonian.

This paper will concentrate on those issues relating to the effects
which are of current interest to experimentalists.  In particular, we
will study BCS-BEC crossover in atomic Fermi gases, looking at a wide
range of different experiments. Our group has been principally
interested in the effects of finite temperature
\cite{Chen1,ourreview,ReviewJLTP} and in this review, we will discuss
how temperature $T$ enters into the standard crossover theory, and how
temperature can be measured, and how temperature changes the character
of the gas from a superfluid at low $T$ to an unusual (but strongly
interacting) normal fluid at moderate $T$, and to ultimately a Fermi gas
at high $T$.  Temperature, of course, plays an important role in these
Fermi gas experiments. As we shall see in this Review, the earliest
evidence for superfluidity is generally based
\cite{Jin3,Grimm,Jin4,Ketterle3,KetterleV,Thomas2,Grimm3,ThermoScience}
on a comparison of the behavior of the gas at two different
temperatures, presumably, one above and one below $T_c$.
Experimentally, one is never strictly at $T=0$, and as a result there
are thermal excitations of the gas which need to be characterized both
theoretically and experimentally.

Before we begin with the ultracold gases, it is useful to discuss to
what extent the BCS-BEC crossover scenario relates to the high
temperature copper oxide superconductors \cite{ourreview,ReviewJLTP}.
While there is much controversy in the field of high $T_c$
superconductivity, there is a school of thought
\cite{ourreview,PS05,TDLee1,Deutscher,Uemura,Randeriareview,LeggettNature}
which argues that these systems are somewhere intermediate between BCS
and BEC.  These arguments are based on the observation that $T_c$ is
anomalously high, and therefore the attractive interaction is presumably
stronger than in conventional superconductors. These are
quasi-two-dimensional (2D) materials which means that they have a
tendency to exhibit ``pre-formed" pairs, that is, pairing in advance of
Bose condensation. This is often referred to as fluctuation effects, but
pre-formed pairs are naturally associated with a more BEC-like scenario.
Importantly these pairs can be used to explain the widespread and
anomalous ``pseudogap" effects \cite{Chen2,JS2} which are the focus
\cite{LeeReview} of most of the current attention in the field. The
existence of ``pre-formed" pairs means that a characteristic (pseudogap)
energy must be supplied in the normal state to break the pairs and
create fermionic excitations.  Hence we say that there is a gap or, more
precisely, a pseudogap in the fermionic excitation spectrum.

The case that the crossover scenario is relevant to the cuprates was
made quite eloquently by A.~J. Leggett in a recent status report
\cite{LeggettNature} on high $T_c$ superconductivity. In this article he
summarized the eight salient ``facts" about the cuprates. It is worth
quoting one here, which relates to their anomalously short coherence
length. \textit{ ``The size of the pairs is somewhere in the range 10-30\,\AA\ --
from measurements of the upper critical field, Fermi velocity and
$T_c$. This means that the pair size is only moderately greater than the
inter-conduction electron in-plane spacing, putting us in the
intermediate regime of the so-called Bose-Einstein condensate to BCS
superconductor (BEC-BCS) crossover, and leading us to expect very large
effects of fluctuations (they are indeed found)."}

The field of BCS-BEC crossover is built around early observations by
Eagles \cite{Eagles} and Leggett \cite{Leggett} that the BCS ground
state, proposed by Bardeen, Cooper, and Schrieffer in 1957, is much more
general than was originally thought.  If one increases the strength of
the attraction and self-consistently solves for the fermionic chemical
potential $\mu$, this wave function will correspond to a more BEC-like
form of superfluidity.
Knowing the ground state what is the nature of the superfluidity at
finite $T$? That is the central question we will address in this Review.

Even without a detailed theoretical framework we can make three
important observations.
\begin{itemize}
\item As we go from BCS to BEC, pairs will form above $T_c$ without
  condensation.  In the normal state, it pays in general to take
  advantage of the attractive pairing interaction.  Only in the extreme
  BCS limit do pairs form exactly at $T_c$.
\item The fundamental statistical entities in these superfluids are
  fermions. We can think of pairs of fermions as a form of ``boson",
  although the statistics are not precisely the same.  We measure these
  ``bosonic" or pair-degrees of freedom indirectly via the fermionic gap
  parameter $\Delta(T)$. In the fermionic regime
  this parameter is the minimum
  energy which must be supplied to create fermionic excitations.
  It tells us about bosons indirectly through the
  binding together of two fermions.
\item In general there will be two types of excitations in these BCS-BEC
  crossover systems. Only in strict BCS theory are the excitations of
  solely fermionic character, and only in the strict BEC limit will the
  excitations be solely of bosonic character. More generally in the
  intermediate case (often called the ``unitary" regime) the excitations
  consist of a mix of both fermions and bosons.
\end{itemize}

\begin{figure}
\centerline{\includegraphics[width=3.5in,clip]{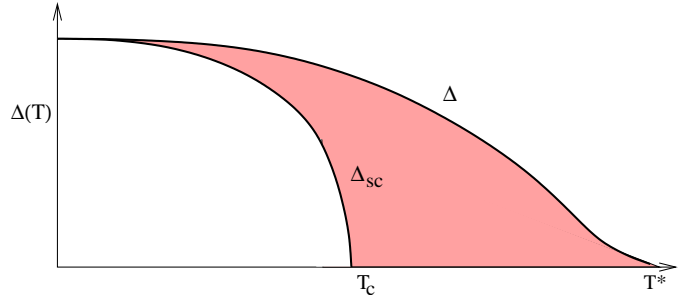}}
\caption{Contrasting behavior of the excitation gap $\Delta(T)$
and superfluid order parameter $\Delta_{sc}(T)$ versus temperature.
The height of the shaded region roughly reflects the density of
noncondensed pairs at each temperature.}
\label{fig:Delta_Deltasc}
\end{figure}

\begin{figure}
\centerline{\includegraphics{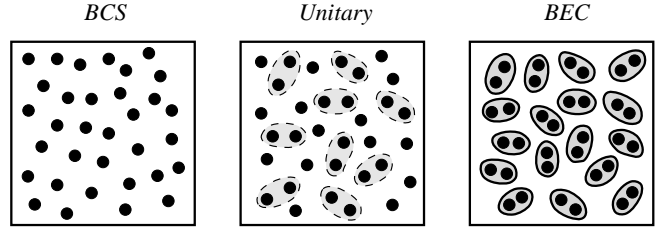}}
\caption{Schematic illustration of excitations in the BCS, unitary and
BEC regimes. The black discs represent fermionic excitations. Pair
excitations become progressively dominant as the system evolves from
the BCS to BEC regime.}
\label{fig:3}
\end{figure}

These observations are illustrated by Figs.~\ref{fig:Delta_Deltasc} and
\ref{fig:3}. In Fig.~\ref{fig:Delta_Deltasc} we schematically plot the
gap parameter $\Delta(T)$ as a function of $T$, along with the
superfluid order parameter $\Delta_{sc}(T)$. The former, which
represents the ``bosonic" degrees of freedom, shows that pairs
continuously form once temperature is less than a crossover temperature
$T^*$, while the order parameter turns on as in a second order phase
transition at $T_c$. The height of the shaded region reflects the number
of noncondensed pairs. This number increases monotonically with
decreasing $T$, until $T_c$ is reached.  A $T$ further decreases below
$T_c$ the number of noncondensed pairs begins to decrease monotonically
due to the condensation of zero momentum pairs.

In Fig.~\ref{fig:3} we present a schematic plot of the excitation type,
which shows that between BCS and BEC (i.e., in the unitary regime) there
will be a mix of fermions and bosons. These bosons and fermions
are not separate fluids, but rather
they are strongly inter-connected. Indeed, the gap in the fermionic
spectrum (related to $\Delta$) is a measure of the number of bosons
in the system.

\begin{figure}
\centerline{\includegraphics[]{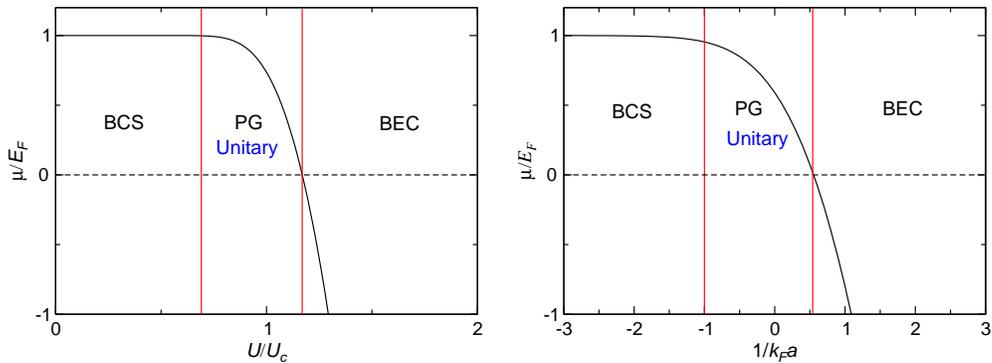}}
\caption{Typical behavior of the chemical potential $\mu$ at $T=0$ in
  the three regimes, as a function of the interaction strength $U/U_c$,
  or, equivalently, $1/k_Fa$. As $U/U_c$ increases from 0, the chemical
  potential $\mu$ starts to decrease and then becomes negative. The
  character of the system changes from fermionic ($\mu>0$) to bosonic
  ($\mu < 0$).  The pseudogap (PG) or unitary regime corresponds to
  non-Fermi liquid based superconductivity, and $U_c (<0)$ corresponds
  to critical coupling for forming a two fermion bound state in vacuum
  \cite{ourreview}, i.e., the unitary point where the two-body $s$-wave
  scattering length $a$ diverges.}
\label{fig:2a}
\end{figure}

\section{Theory Outline}
In this paper we will presume the ground state is given by the standard
BCS-like wavefunction,
\begin{equation}
\Psi_0=\Pi_{\bf k}(\uk+\vk c_{\bf k,\uparrow}^{\dagger}
c_{\bf -k,\downarrow}^{\dagger})|0\rangle \,,
\label{eq:1a}
\end{equation}
where $ c^\dag_{\bf k,\sigma}$ and $ c^{}_{\bf k,\sigma}$ are the
creation and annihilation operators for fermions of momentum ${\bf k}$
and spin $\sigma=\uparrow,\downarrow$.  The variational parameters $\vk$
and $u_{\bf k}$ are usually represented by the two more directly
accessible parameters $\Delta_{sc}(T=0)$ and $\mu$, which characterize
the fermionic system. Here $\Delta_{sc}(T=0)$ is the zero temperature
superconducting order parameter. These fermionic parameters are uniquely
determined in terms of the attractive interaction $U$ and the fermionic
density $n$. The variationally determined self consistency conditions
are given by two BCS-like equations which we refer to as the ``gap" and
``number" equations, respectively.

At $T=0$, the effects of BEC-BCS crossover are most directly reflected
in the behavior of the fermionic chemical potential $\mu$, which is
associated with the number equation.  We plot the behavior of $\mu$ in
Fig.~\ref{fig:2a}, which indicates the BCS and BEC regimes.  In the weak
coupling regime $\mu = E_F$ and ordinary BCS theory results.  
With increasing $|U|$, 
$\mu$ begins to decrease, eventually
crossing zero and then ultimately becoming negative in the BEC regime.
We generally view $\mu = 0 $ as a crossing
point.  For positive $\mu$ the system has a remnant of a Fermi surface,
and we say that it is ``fermionic".  For negative $\mu$, the Fermi
surface is gone and the material is ``bosonic".

One can debate whether other ground states ought to be considered.
Indeed the work of the Camerino group \cite{Strinati2,Strinati5} is
based on a finite $T$ approach first introduced by Nozieres and
Schmitt-Rink (NSR) \cite{NSR,Randeriareview}.  This leads to a
different, and not as readily characterized ground state.  We list some
arguments in support of the ground state in Eq.~(\ref{eq:1a}).  (i) This
is the basis for the widely studied Bogoliubov-de Gennes approach, which
can be applied \cite{our_vortex,Machida,Ho_vortex,Kinnunen} to the
BCS-BEC crossover problem at $T=0$.  (ii) At arbitrarily strong coupling
(and $T=0$), this ground state can be shown to coincide with a
Gross-Pitaevskii (GP) description \cite{PSP03} of the boson system.
(iii) This ground state is the basis for the rapidly proliferating
theoretical literature \cite{Kinnunen,SR06,SM06,HS06, Tsinghuagroup,
PWY05,PS05a} on spin polarized Fermi superfluids.  In addition to our
own work there have been some additional studies which include the
effects of temperature \cite{Machida2,YD05,LatestStoof}, albeit at
a lower order mean field theory than considered here.

\begin{table}
\caption{BCS theory by way of BEC analogy.  Here we compare condensation
in composite and point bosons; $\mu_B$ is the bosonic chemical
potential, $N_0$ and $N_T$ are the number of condensed and noncondensed
bosons, respectively. We define $\mu_{pair}$ as the chemical potential
for the noncondensed pairs.  Here $\Delta(T)$ is the total fermionic gap
which contains contributions from the noncondensed ($\Delta_{pg}^2$) and
condensed terms ($\Delta_{sc}^2$). In the strict BCS limit $\Delta_{pg}
=0$, so that the order parameter and gap are identical. }
\begin{tabular}{|p{1.35in}|p{2.05in}|p{1.3in}|}
\hline
& \parbox[c][5mm][c]{2.in}{\centering{Composite bosons}} &
\parbox[c][5mm][c]{1.3in}{\centering{Point bosons}} \\
\hline
\parbox[c][5mm][c]{1.3in}{\centering{Pair chemical potential}} &
{{\parbox[c][8mm][c]{2.in}{\centering{$\mu_{pair}=0$,
$T\leq T_c$} \\ Leads to BCS gap equation for $\Delta(T)$}}} &
\parbox[c][5mm][c]{1.3in}{\centering{$\mu_{B}=0$, $T\leq T_c$}} \\
\hline
\parbox[c][5mm][c]{1.3in}{\centering{Total ``number'' of pairs}} &
\parbox[c][5mm][c]{2.in}
{\centering{$\Delta^{2}(T)=\Delta_{sc}^{2}(T)+\Delta_{pg}^{2}(T)$}}
& \parbox[c][5mm][c]{1.3in}{\centering{$N=N_{0}+N_{T}$}} \\
\hline
\parbox[c][5mm][c]{1.3in}{\centering{Noncondensed pairs}} &
\parbox[c][5mm][c]{2.in}{\centering{$Z\Delta_{pg}^{2}=\sum_{\mathbf{q}\neq
0}b(\Omega_{q})$}} &
\parbox[c][5mm][c]{1.3in}{\centering{$N_{T}=\sum_{\mathbf{q}\neq
      0}b(\Omega_{q})$}} \\ 
\hline
\end{tabular}
\end{table}

We begin at the more physical level by stressing the analogy between
condensation in this composite boson or fermionic superfluid and
condensation in a gas of ideal point bosons. Our microscopic theory
treats self-consistently two-particle and one-particle Green's functions
on an equal footing.  Because the physics is so simple and clear, we can
fairly readily anticipate the form of the central equations of this
BCS-BEC generalization of BCS theory. It is important to stress,
however, that these equations can be derived more rigorously from a
truncated series of equations of motion for the appropriate Green's
functions \cite{Chen2,ChenThesis}.

There are three principal equations which govern Bose condensation: the
vanishing of the bosonic chemical potential at all $ T \leq T_c$ is the
first.  We will refer to this condition as the ``BEC condition".  It is
related to the usual Thouless criterion for superconductivity, but the
latter is generally associated only with the temperature $T_c$. The
second equation is the boson number equation.  All ``bosons" must be
accounted for as either condensed or noncondensed.  The third equation
is the number of noncondensed ``bosons", which are created by thermal
excitations. This is determined simply by inserting the known excitation
spectrum of the excited pairs or bosons, into the Bose distribution
function $b(x)$.  With this equation, and the second equation, one can
then deduce the number of condensed bosons.

These three central equations for bosons are indicated in Table I, on
the far right, for true point bosons, and in the second column for the
composite bosons which appear in fermionic superfluids.  For these
composite bosons the quantity which provides a measure of the ``number"
of bosons ($N$) is given by $\Delta^2(T)$ (up to a constant coefficient,
$Z$).  How does one quantitatively establish the appropriate ``boson
number" for the fermionic case?  This is determined via the self
consistent gap equation for $\Delta(T)$, which, in turn, is determined
using the first condition: that the pair chemical potential $\mu_{pair}$
is zero at and below $T_c$.  How does one compute the number of excited
pairs?  Once the gap equation is interpreted in terms of the appropriate
noncondensed pair propagator (see below), then one knows the related
excitation spectrum $\Omega_q$ of this propagator.

The quantity $Z$ which appears in the last equation of the Table (for
the composite bosons) gives the relation between the gap associated with
noncondensed pairs ($\Delta_{pg}^2$) and the number of all pairs [$\sum
b(\Omega_q)$]. It can be readily calculated in this theory; once one has
the noncondensed pair propagator, $Z$ appears as the inverse residue.
(Deep in the BEC regime \cite{JS3}, $Z$ is relatively simple to compute,
for here the boson number density approaches the asymptote $n/2$).  More
precisely, the total number of bosons in the present case has to be
determined self-consistently through the gap equation for $\Delta(T)$.
It also involves the fermion number equation through the related
fermionic chemical potential.

To be consistent with the ground state variational conditions, the
vanishing of the pair chemical potential is associated with a particular
form for the pair propagator involving dressed Green's functions.
These, in turn, determine the fermionic chemical potential through the
fermion number equation.

\subsection{Microscopic T-matrix Scheme}

Next, we implement this picture microscopically via a $T$-matrix
approximation. We include spin indices throughout so that it will be
clear how to apply this scheme to spin polarized superfluids.  This
means that we consider the coupled equations between the particles (with
propagator $G$) and the pairs [with propagator $t(P)$] and drop all
higher order terms.  This theory does not include direct ``boson-boson"
interactions, although the pairs do interact indirectly via the
fermions, in an averaged or mean field sense.
Here, for all $T\leq T_{c}$, the BEC condition is interpreted as
requiring that the pair chemical potential $\mu_{pair}$ associated with
the noncondensed pairs, vanish.  Within a $T$-matrix scheme, the pair
propagator is given by
\begin{equation}
t_{pg}^{-1}(P)=U^{-1}+\chi(P)\,,
\label{eq:1}
\end{equation}
where $\chi$ is the pair susceptibility.  The function $\chi(P)$ is, in
many ways, the most fundamental quantity we introduce.
We will show that one obtains consistent answers between $T$-matrix
based approaches and the BCS-Leggett ground state equations,
provided the components of the pair susceptibility, defined by
\begin{equation}
\chi(P)=\frac{1}{2}\big[\chi_{\uparrow\downarrow}(P)+
\chi_{\downarrow\uparrow}(P)\big]
\label{eq:2}
\end{equation}
are given by the product of one dressed and one bare Green's function
\begin{equation}
\label{eq:3}
\hspace*{1cm}\chi_{\uparrow\downarrow}(P)=\sum_{K}G_{0\uparrow}(P-K)G_{\downarrow}(K),  \qquad
\chi_{\downarrow\uparrow}(P)=\sum_{K}G_{0\downarrow}(P-K)G_{\uparrow}(K)\,,
\end{equation}
where $P=(i\Omega_l,\mathbf{p})$, and $G$ and $G_0$ are the full and
bare Green's functions respectively. 
Here $G_{0,\sigma}^{-1} (K) = i \omega_{n} - \xi_{\mathbf{k},\sigma}$,
$\xi_{\mathbf{k},\sigma} = \ek-\mu_\sigma$, $\ek=\hbar^2k^2/2m$ is the
kinetic energy of fermions, and $\mu_\sigma$ is the fermionic chemical
potential for spin $\sigma=\uparrow,\downarrow$.  Throughout this paper,
we take $\hbar=1$, $k_B=1$, and use the four-vector notation $K\equiv
(i\omega_n, \mathbf{k})$, $P\equiv (i\Omega_l, \mathbf{q})$, $\sum_K
\equiv T\sum_n \sum_{\bf k}$, etc, where $\omega_n = (2n+1)\pi T$ and
$\Omega_l = 2l\pi T$ are the standard odd and even Matsubara frequencies
\cite{Fetter} (where $n$ and $l$ are integers).

We now evaluate the BEC condition
\begin{equation}
t_{pg}^{-1} (0) = 0 = U^{-1}+\chi(0)\,.
\label{eq:5}
\end{equation}
The one-particle Green's function for fermions with spin $\sigma$ is
\begin{eqnarray}
G^{-1}_{\sigma}(K)&=&G^{-1}_{0\sigma}(K)-\Sigma_{\sigma}(K) 
= i\omega_{n} -\xi_{k\sigma}-\Sigma_{\sigma}(K) \,,
\label{eq:13}
\end{eqnarray}
where $\bar{\sigma}\equiv -\sigma$, and the self-energy $\Sigma_{\sigma}$ is
of the BCS-like form
\begin{equation}
\Sigma_{\sigma}(K)=-\Delta^{2}G_{0\bar{\sigma}}(-K) =
\frac{\Delta^{2}}{i\omega+\xi_{k\bar{\sigma}}} \,.
\label{eq:14}
\end{equation}
It should be noted that we use a contact potential so that the symmetry
factor $\phik$ associated with the pairing interaction is trivially 
$\phik = 1$. For a nontrivial $\phik$, one only needs to replace
$\Delta$ with $\Delta\phik$ in Eq.~(\ref{eq:14}).  We will see below how
this form for the self energy very naturally arises (below $T_c$) in a
$T$-matrix approach.  Thus
\begin{equation}
G^{-1}_{\sigma}(K)= i\omega-\xi_{k\sigma}-\frac{\Delta^{2}}
{i\omega+\xi_{k\bar{\sigma}}}\,.
\label{eq:15}
\end{equation}
 
Now we are in position to calculate the pair susceptibility at $P=0$ 
%
\begin{eqnarray}
\chi(0) &=& \chi_{\uparrow\downarrow}(0) =
\chi_{\downarrow\uparrow}(0)
=-\sum_{K}\frac{1}{(i\omega_{n}-E_{k\downarrow})(i\omega_{n}+E_{k\uparrow})}
\,.
\label{eq:17}
\end{eqnarray}
Substituting this expression into our BEC condition Eq.~(\ref{eq:5}), we
obtain the gap equation 
\begin{eqnarray}
0
&=&\frac{1}{U}+\sum_{\mathbf{k}}
\left[\frac{1-f(E_{k\downarrow})-f(E_{k\uparrow})}{2E_{k}}
\right]
= \frac{1}{U}+\sum_{\mathbf{k}}\frac{1-2\bar{f}(E_{k})}{2E_{k}} \,.
\label{eq:10}
\end{eqnarray}
after carrying out the Matsubara summation.
Here $\mu=(\mu_{\uparrow}+\mu_{\downarrow})/2$ and
$h=(\mu_{\uparrow} -\mu_{\downarrow})/2$,
$\Ek = \sqrt{\xik^2 +\Delta^2}$, $E_{k\uparrow}=-h+E_{k}$ and
$E_{k\downarrow}=h+E_{k}$, where $\xi_{k}=\epsilon_{k}-\mu$.
In addition, we define the average
$ \bar{f}(x) \equiv [f(x+h)+f(x-h)]/2, $
where $f(x)$ is the Fermi distribution function. 


The coupling constant $U$ can be replaced in favor of the dimensionless
parameter, $1/k_Fa$, via the relationship $m/(4\pi a) = 1/U +
\sum_{\mathbf{k}}(2\epsilon_{k})^{-1}$, where $a$ is the two-body
$s$-wave scattering length, and $k_F$ is the noninteracting Fermi wave
vector for the same total number density in the absence of population
imbalance.
Therefore the gap equation can be rewritten as
\begin{equation}
-\frac{m}{4\pi a}=\sum_{\mathbf{k}}\left[\frac{1-2\bar{f}(E_{k})}
  {2E_{k}}-\frac{1}{2\epsilon_{k}} \right] \,.
\label{eq:11}
\end{equation}
Here the ``unitary scattering" limit corresponds to resonant scattering
where $ a \rightarrow \infty$.  This scattering length is tunable by
magnetic field application and we say that we are on the BCS or BEC side
of resonance, depending on whether the fields are higher or lower than
the resonant field, or alternatively whether $a$ is negative or
positive, respectively.

Finally, in terms of Green's functions, we readily arrive at the number
equations: $n_{\sigma}=\sum_{K}G_{\sigma}(K)$, which are consistent
with their ground state counterparts
\begin{eqnarray}
n_{\sigma}&=&\sum_{\mathbf{k}}[f(E_{k\sigma})u_{\mathbf{k}}^{2}+
f(E_{k\bar{\sigma}}) v_{\mathbf{k}}^{2}] \,,
\label{eq:12}
\end{eqnarray}
where 
the coherence factors $u_\mathbf{k}^2, v_\mathbf{k}^2 = (1\pm
\xi_\mathbf{k}/\Ek)/2$.
  
\begin{figure}
\centerline{\includegraphics[]{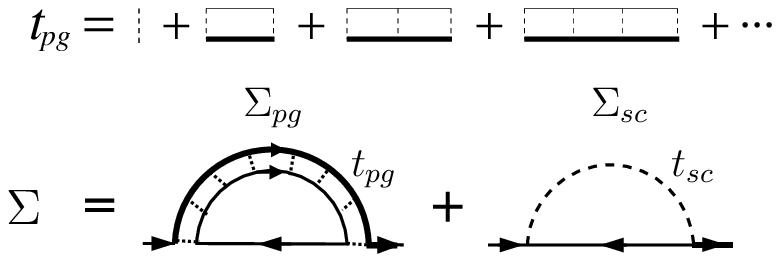}}
\caption{$T$-matrix and self-energy diagrams for the present $T$-matrix
  scheme. The self-energy comes from contributions of both condensed
  ($\Sigma_{sc}$) and noncondensed ($\Sigma_{pg}$) pairs. Note that
  there is one dressed and full Green's function in the $T$-matrix. Here
  $t_{pg}$ represents the propagator for the noncondensed pairs.}
\label{fig:T-matrix}
\end{figure}

Next we use this $T$-matrix scheme to derive Eq.~(\ref{eq:14}) and
separate the contribution from condensed and noncondensed pairs.  The
diagrammatic representation of our $T$-matrix scheme is shown in
Fig.~\ref{fig:T-matrix}.  The first line indicates $t_{pg}$, and the
second the total self energy.  One can see throughout the combination of
one dressed and one bare Green's function, as represented by the thick
and thin lines.
The self energy consists of two contributions from the noncondensed
pairs or pseudogap ($pg$) and from the condensate ($sc$).  There are,
analogously, two contributions in the full $T$-matrix
\begin{eqnarray}
t &=& t_{pg} + t_{sc} \,, \label{t-matrix}\\
t_{pg}(P)&=& \frac{U}{1+U \chi(P)}, \qquad P \neq 0 \,,
\label{t-matrix_pg}\\
t_{sc}(P)&=& -\frac{\Delta_{sc}^2}{T} \delta(P) \,,
\label{t-matrix_sc}
\label{eq:43}
\end{eqnarray}
where we write
$\Delta_
{sc}=-U \sum _{\bf k}\langle c_{-{\bf k}\downarrow}c_{{\bf
k}\uparrow}\rangle$.

Similarly, we have for the fermion self energy
\begin{equation}
\Sigma_\sigma (K) = \Sigma_\sigma ^{sc}(K) + \Sigma_\sigma ^ {pg} (K) =
\sum_P t(P) G_{0,\bar{\sigma}} (P-K) \,. 
\label{eq:sigma2}
\end{equation}
We can see at once that
\begin{equation}
\Sigma_\sigma ^{sc}(K) = \sum_P t_{sc}(P) G_{0,\bar{\sigma}}(P-K) =
-G_{0,\bar{\sigma}} (-K) \Delta_{sc}^2
\,.
\label{eq:70}
\end{equation}
The vanishing of the pair chemical potential implies that
\begin{equation}
t_{pg}^{-1} (0) = U^{-1} + \chi(0) = 0, \qquad T \leq T_c \;.
\label{eq:3e}
\end{equation}
Moreover, a vanishing chemical potential means that $t_{pg}(P)$
diverges $P=0$. Thus, we may approximate \cite{Maly1}
Eq.~(\ref{eq:sigma2}) to yield
\begin{equation}
\Sigma_{\sigma} (K)\approx -G_{0,\bar{\sigma}} (-K) \Delta^2 \,,
\label{eq:sigma3}
\end{equation}
where
\begin{equation}
\Delta^2 (T) \equiv \Delta_{sc}^2(T)  + \Delta_{pg}^2(T) \,,
\label{eq:sum}
\end{equation}
Importantly, we are led to identify the quantity $\Delta_{pg}$
\begin{equation}
\Delta_{pg}^2 \equiv -\sum_{P\neq 0} t_{pg}(P).
\label{eq:delta_pg}
\end{equation}
Note that in the normal state (where $\mu_{pair}$ is non-zero)
Eq. (\ref{eq:sigma3}) is no longer a good approximation.

We now have a closed set of equations for addressing the ordered phase.
This approach can be readily generalized \cite{LOFFlong} to treat more exotic
polarized phases such as the LOFF state \cite{FFLO,Combescot,LOFF_Review}.
We can similarly extend this approach to temperatures somewhat above
$T_c$, by self consistently including a non-zero pair chemical
potential.  This is a necessary step in addressing a trap as well
\cite{ChienRapid}.  Additionally, the propagator for noncondensed pairs
can now be quantified, using the self consistently determined pair
susceptibility.
At small four-vector $P$, we may expand the inverse of $t_{pg}$,
after analytical continuation ($i\Omega_l \rightarrow \Omega+i0^+$), to
obtain
\begin{equation}
t_{pg}^{-1} \approx a_1\Omega^2 + Z\left(\Omega - \frac{p^2}{2 M^*} + \mu
_{pair} +i \Gamma^{}_P\right),
\label{Omega_q:exp}
\end{equation}
where the imaginary part $\Gamma^{}_P \rightarrow 0$ rapidly as
$p\rightarrow 0$ below $T_c$.  Because we are interested in the moderate
and strong coupling cases, we drop the $a_1 \Omega^2$ term in
Eq. (\ref{Omega_q:exp}), and hence
\begin{equation}
t_{pg}(P) = \frac { Z^{-1}}{\Omega - \Omega^{}_p +\mu_{pair} + i \Gamma^{}_P},
\label{eq:expandt}
\end{equation}
where we associate
\begin{equation}
\Omega_{\mathbf{p}} \approx \frac{p^2} {2 M^*} \,.
\label{eq:53}
\end{equation}
  This establishes a quadratic pair dispersion and defines the effective
pair mass, $M^*$. They can be calculated via a small $p$ expansion of
$\chi(P)$,
\begin{equation}
Z=\left.\frac{\partial \chi}{\partial\Omega}\right|_{\Omega=0,\mathbf{p}=0},
\qquad 
\frac{1}{2M^*} =-\left.\frac{1}{6Z}\frac{\partial^{2} \chi}{\partial
p^{2}}\right|_{\Omega=0,\mathbf{p}=0} \,.
\end{equation}
Finally, one can rewrite Eq. (\ref{eq:delta_pg}) as
\begin{equation}
\Delta_{pg}^2 (T) = Z^{-1}\sum_{\mathbf{p}} b(\Omega_p) \,.
\label{eq:81}
\end{equation}

We now return to the strong analogies between our $T$-matrix approach
and Bose condensation of point bosons, as summarized in Table I.
We have three central equations.

1. The pair chemical potential must vanish at and below $T_c$,
\begin{equation}
\mu_{pair} = 0, \quad (T \leq T_c).
\label{eq:57}
\end{equation}
Importantly this condition leads to the mean field gap equation derived
in Eq.~(\ref{eq:10}).

2. There must be a conservation of the total number of (composite)
``bosons" in the system.  For this condition, our central equation is
Eq.~(\ref{eq:sum}).  Here it is understood that the number of "bosons"
is effectively represented by the parameter $\Delta^2 (T)$.  Unlike the
point boson case, the ``total boson number" is temperature dependent and
has to be self-consistently determined. 

3. The number of noncondensed pairs is readily computed in terms of the
pair dispersion, just as in conventional BEC. For this condition our
central equation is Eq.~(\ref{eq:81}).  Then, just as in conventional
BEC, the number of condensed bosons (proportional to $\Delta_{sc}^2$) is
determined by the difference between $\Delta^2(T)$ and $\Delta_{pg}^2
(T)$.  This, in turn, determines $T_c$ as the lowest temperature(s) in
the normal state at which noncondensed pairs exhaust the total weight of
$\Delta^2$ so that $\Delta_{pg}^2 = \Delta^2$.  Solving for the
``transition temperature" in the absence of pseudogap effects
\cite{Machida2,YD05,LatestStoof} leads to the quantity $T_c^{MF}$.  More
precisely, $T_c^{MF}$ should be thought of as the temperature at which
the excitation gap $\Delta(T)$ vanishes.  This provides a reasonable
estimate, for the pairing onset temperature $T^*$, (when a stable
superfluid phase exists).  This is distinguished from the transition
temperature. We note that $T^*$ represents a smooth crossover rather
than an abrupt phase transition.

It should be stressed that the dispersion relation for the noncondensed
pairs is quadratic.  While \textit{one will always find a linear
dispersion in the collective mode spectrum \cite{Kosztin2}}, within the
present class of BCS-BEC crossover theories, the restriction to a
$T$-matrix scheme means that there is no feedback from the collective
modes onto the pair excitation spectrum.  In effect, the $T$-matrix
approximation does not incorporate pair-pair interactions at a level
needed to arrive at this expected linear dispersion in the pair
excitation spectrum.  Nevertheless, this level of approximation
is consistent with the underlying ground state wavefunction.

\section{Behavior of $T_c$ and Trap Effects}

\begin{figure}
\centerline{\includegraphics[]{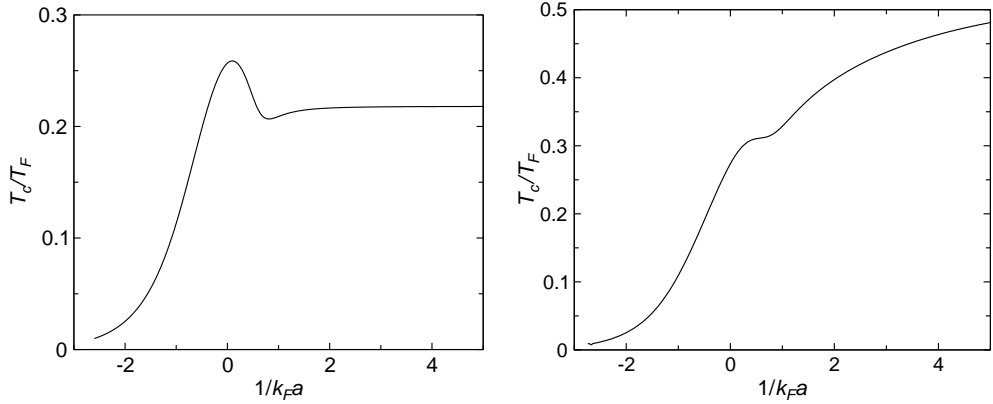}}
\caption{Typical behavior of $T_c$ as a function of $1/k_Fa$ in a
  homogeneous system (left panel) and in a trapped Fermi gas (right
  panel). $T_c$ follows the BCS predictions and approaches the BEC
  asymptote $0.218T_F$ and $0.518T_F$ in the homogeneous and trapped
  cases, respectively. In contrast to the homogeneous case, the BEC
  asymptote in a trap is much higher due to a compressed profile for
  trapped bosons. In the homogeneous case, $T_c$ reaches a maximum
  around $1/k_Fa=0$ and a minimum around where $\mu=0$. In the trapped
  case, this maximum/minimum behavior is washed out largely by the
  shrinking cloud size as $1/k_Fa$ increases.}
\label{fig:Tc}
\end{figure}

Before turning to experiment, it is important to discuss the behavior of
$T_c$ which is plotted as a function of scattering length in the left
panel of Fig.~\ref{fig:Tc} for the homogeneous case, presuming $s$-wave
pairing.  Starting from the BCS regime this figure shows that $T_c$
initially increases as the interaction strength increases. However, this
increase competes with the opening of a pseudogap or excitation gap
$\Delta(T_c)$. Technically, the pairs become effectively heavier before
they form true bound states.  Eventually $T_c$ reaches a maximum (very
near unitarity) and then decreases slightly until field strengths
corresponding to the point where $\mu$ becomes zero.  At this field
value (essentially where $T_c$ is minimum), the system becomes a
``bosonic" superfluid, and beyond this point $T_c$ increases slightly to
reach the asymptote $0.218E_F$ corresponding to an ideal Bose gas.

Trap effects change these results only quantitatively as seen in the
right panel of Fig.~\ref{fig:Tc}. Here the maximum in $T_c$ may no
longer be visible. The calculated value of $T_c$ ($\sim 0.3 T_F$) at
unitarity is in good agreement with experiment
\cite{ThermoScience,Thomasnew} and other theoretical estimates
\cite{Strinati4}. To treat these trap effects one introduces the local
density approximation (LDA) in which $T_c$ is computed under the
presumption that the chemical potential $\mu \rightarrow \mu - V(r)$ .
Here we consider a spherical trap with $V(r)=\frac{1}{2}m\omega^2 r^2$.
The Fermi energy $E_F$ is determined by the total atom number $N$ via
$E_F \equiv k_BT_F = \hbar\omega (3N)^{1/3} \equiv \hbar^2k_F^2/2m$,
where $k_F$ is the Fermi wavevector at the center of the trap. It can be
seen that the homogeneous curve is effectively multiplied by an
``envelope" curve when a trap is present.  This envelope, with a higher
BEC asymptote, reflects the fact that the particle density at the center
of the trap is higher in the bosonic, relative to the fermionic case. In
this way $T_c$ is relatively higher in the BEC regime, as compared to
its counterpart in the homogeneous case.

\begin{figure}
\centerline{\includegraphics[width=2.8in,clip]{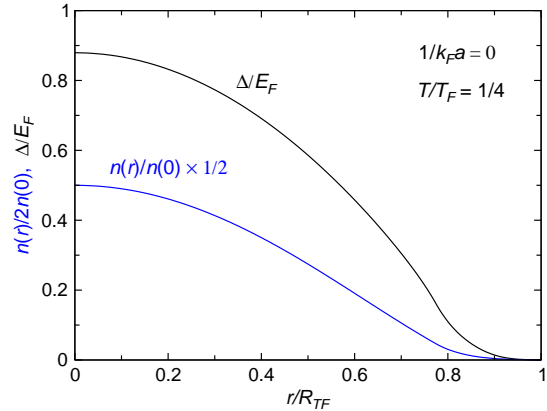}}
\caption{Typical spatial profile for density $n(r)$ and fermionic
  excitation gap $\Delta(r)$ of a Fermi gas in a trap. The curves are
  computed at $T=T_F/4$ and at unitarity, where $1/k_Fa=0$. Here $R_{TF}$ is
  the Thomas-Fermi radius, which gives the cloud size at $T=0$ in the
  noninteracting limit.}
\label{fig:denstrap}
\end{figure}

Figure \ref{fig:denstrap} is a central one, for it prepares us for
understanding various experiments.  It presents a plot of the position
dependent excitation gap $\Delta(r)$ and particle density $n(r)$ profile
over the extent of the trap.  An important point needs to be made:
because the gap is largest at the center of the trap, bosonic
excitations will be dominant there. At the edge of the trap, by
contrast, where fermions are only weakly bound (since $\Delta(r)$ is
small), the excitations will be primarily fermionic.  We will see the
implications of these observations as we examine thermodynamic
\cite{ChenThermo} and radio frequency (RF) spectra data \cite{heyan} in
the ultracold gases.

One should appreciate that temperature is not straightforward to measure
in these cold gases; it is most readily measured at the Fermi gas
endpoint \cite{Jin4} or in the deep BEC regime \cite{Grimm4}. At
unitarity, the physical temperature can be extracted using
phenomenological fits to the particle density profiles based on the
universality hypothesis \cite{JasonHo,Thomasnew,ThomasVirial} with
proper re-calibration \cite{ThermoScience,JS5} below $T_c$. For a more
general magnetic field one has to resort to adiabatic sweep thermometry
\cite{ChenThermo,Williams}.  Here, the magnetic field of interest is
accessed via a slow, adiabatic, or isentropic, magnetic field sweep
starting from either the BCS or BEC endpoints, where the temperature
(and the entropy \cite{Carr}) are known.  A finite temperature theory of
BCS-BEC crossover is required to calculate the entropy \cite{ChenThermo}
at general magnetic fields.  In this way, the physical temperature can
be associated with the measured endpoint temperature.  Indeed, we will
see below that the temperature which appears in the measured superfluid
phase diagram \cite{Jin4} or in the RF pairing gap experiments
\cite{Grimm4} is given in terms of the endpoint temperature.

\section{Experimental Evidence for a Pseudogap in Cold Gases}

Our finite temperature generalization of the BCS-like ground state has
introduced the concept of a ``pseudogap".  This pseudogap in the
fermionic spectrum should be viewed as synonymous with the concept of
noncondensed pairs, or with pairs which have a finite center of mass
momentum.  They are important both above and below $T_c$. In this
section we want to explore the evidence for these noncondensed pairs
using three different experiments: density profiles, normal state
thermodynamics and RF pairing gap spectroscopy. In this section we will
consider the case of an unpolarized gas.

\begin{figure}
\centerline{\includegraphics[width=4.5in,clip]{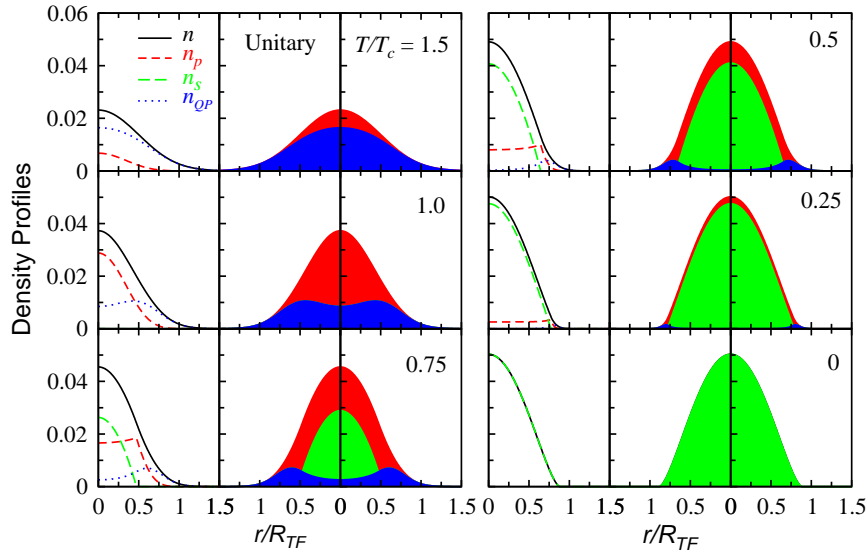}}
\caption{Decomposition of density profiles at various temperatures at
unitarity. Here green (light gray) refers to the condensate, red (dark
gray) to the noncondensed pairs and blue (black) to the excited
fermionic states. $T_c = 0.27T_F$, and $R_{TF}$ is the Thomas-Fermi
radius. The presence of noncondensed pairs is essential \cite{JS5} for
explaining why there are no sharp features in these profiles, associated
with the interface of the normal and superfluid regions. Here $n_s$,
$n_p$, and $n_{QP}$ denote density of superfluid, incoherent pairs,
and fermioinic quasiparticle, respectively.}
\label{fig:23new}
\end{figure}

In Fig. \ref{fig:23new} we plot a decomposition of the particle density
profiles \cite{JS5} for various temperatures above and below $T_c$.  The
various color codes (or gray scales) indicate the condensate along with
the noncondensed pairs and the fermions. This decomposition is based on
the superfluid density so that all atoms participate in the condensation
at $T=0$.

An important observation should be noted.  The noncondensed pairs are
responsible for smoothing out what otherwise would be a discontinuity
between the fermionic and condensate contributions.  This leads to a
featureless profile, in agreement with experiment \cite{Thomas,Grimm2}.
Indeed, these experimental observations presented a challenge for
previous theories \cite{Chiofalo,JasonHo} which ignored noncondensed
pairs, and therefore predicted an effectively bimodal profile with a
kink at the edge of the superfluid core. One can see from the figure
that even at $T_c$, the system is different from a Fermi gas.  That is,
noncondensed pairs are present in the central region of the trap when
the condensate is gone.  Even at $T/T_c = 1.5$ there is a considerable
fraction of noncondensed pairs.
 It is not until around $T^* = 2T_c$ for this unitary case, that 
noncondensed pairs have finally disappeared.

\begin{figure}
\centerline{\includegraphics[width=3.5in,clip]{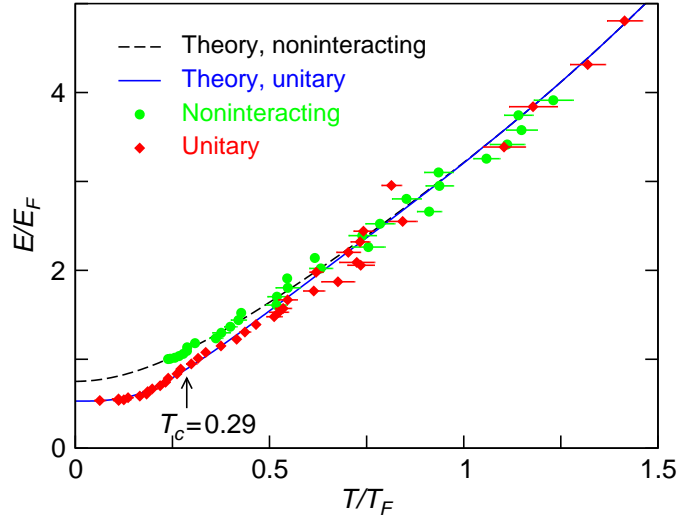}}
\caption{Pseudogap effects as apparent from thermodynamics.  From Ref.
  \cite{ThermoScience}. The fact that the experimental data (symbols)
(and the two theoretical curves) for noninteracting and unitary Fermi gases 
  do not merge until higher $T^* > T_c$ is consistent with the
  presence of a normal state pseudogap.}
\label{fig:232new}
\end{figure}

We next turn to a detailed comparison of theory and experiment for
thermodynamics.  Figure \ref{fig:232new} presents a plot of energy $E$
as a function of $T$ comparing the unitary and non-interacting regimes.
The solid curves are theoretical while the data points are measured in
$^6$Li \cite{ThermoScience}.  There has been a recalibration of the
experimental temperature scale in order to plot theory and experiment in
the same figure.  The latter was determined via Thomas-Fermi fits to the
density profiles.  To arrive at the calibration, we applied the same
fits to the theoretically produced density profiles, examples of which
appear in Fig.~\ref{fig:23new}.  Good agreement between theory and
experiment is apparent in Fig.~\ref{fig:232new}.  In the figure, the
temperature dependence of $E$ reflects primarily fermionic excitations
at the edge of the trap, although there is a small bosonic contribution
as well.
Importantly one can see the effect of a pseudogap in the unitary case.
The temperature $T^*$ is visible from the plots as that at which the
non-interacting and unitary curves merge.  This corresponds roughly
to $T^* \approx 2 T_c$.

\begin{figure}
\centerline{\includegraphics[width=5.in,clip]{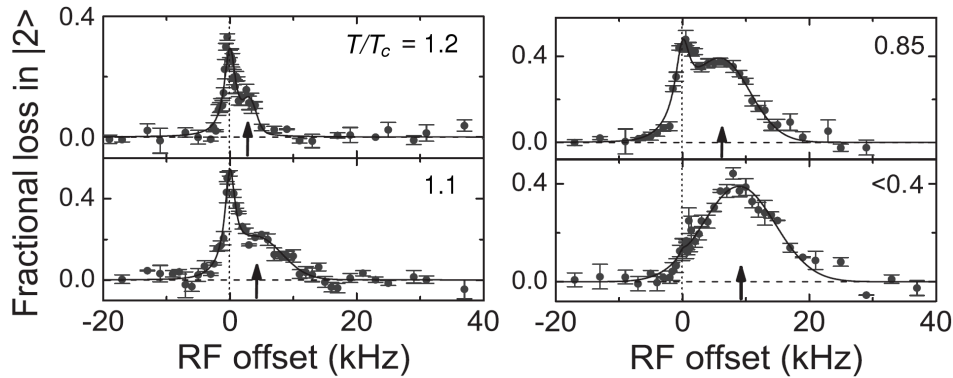}}
\caption{Experimental RF Spectra for $^6$Li at unitarity at 834\,G. The
temperatures labeled in the figure were computed theoretically at
unitarity based on adiabatic sweeps from BEC.  The two top curves, thus,
correspond to the normal phase, thereby, indicating pseudogap
effects. Here $E_F = 2.5 \mu $K, or 52\,kHz.  From Ref.~\cite{Grimm4}.}
\label{fig:24Cheng}
\end{figure}

Measurements \cite{Grimm4} of the excitation gap $\Delta$ can be made
more directly, and, in this way one can further probe the existence of a
pseudogap.  This pairing gap spectroscopy is based on using a third
atomic level, called $|3 \rangle$, which does not participate in the
superfluid pairing. Under application of RF fields, one component of the
Cooper pairs, called $|2 \rangle$, is excited to state $|3\rangle$.  If
there is no gap $\Delta$ then the energy it takes to excite $|2 \rangle$
to $|3 \rangle$ is the atomic level splitting $\omega_{23}$. In the
presence of pairing (either above or below $T_c$) an extra energy
$\Delta$ must be input to excite the state $|2 \rangle$, as a result of
the breaking of the pairs.  Figure \ref{fig:24Cheng} shows a plot of the
spectra for $^6$Li near unitarity for four different temperatures, which
we discuss in more detail below.  In general for this case, as well as
for the BCS and BEC limits, there are two peak structures which appear
in the data and in the theory \cite{Torma2,heyan}: the sharp peak at
$\omega_{23} \equiv 0$ which is associated with ``free" fermions at the
trap edge and the broader peak which reflects the presence of paired
atoms; more precisely, this broad peak derives from the distribution of
$\Delta$ in the trap.  At high $T$ (compared to $\Delta$), only the
sharp feature is present, whereas at low $T$ only the broad feature
remains.  The sharpness of the free atom peak can be understood as
coming from a large phase space contribution associated with the $2
\rightarrow 3$ excitations \cite{heyan}.  These data alone do
not directly indicate the presence of superfluidity, but rather they
provide strong evidence for pairing.

It is interesting to return to discuss the temperatures in the various
panels. What is measured experimentally are temperatures $T'$ which
correspond to the temperature at the start of an adiabatic sweep from
the BEC limit to unitarity. Here fits to the BEC-like profiles are used
to deduce $T'$ from the shape of the Gaussian tails in the trap. Based
on knowledge \cite{ChenThermo} about thermodynamics (energy $E$ in the
previous figure or, equivalently, entropy $S$), and given $T'$, one can
then compute the final temperature in the unitary regime, assuming $S$
is constant.  Indeed, this adiabaticity has been confirmed
experimentally in related work \cite{Grimm2}. We find that the four
temperatures are as indicated in the figures.  Importantly, one can
conclude that the first two cases correspond to a normal state, albeit
not far above $T_c$.  In this way, these figures suggest that a pseudogap is
present as reflected by the broad shoulder above the narrow free atom
peak.

\section{Establishing Superfluidity in Cold Fermi Gases}

From the time of the earliest discoveries
\cite{Kinast2,Jin3,Grimm,Jin4,Ketterle3,KetterleV,Thomas2,Grimm3,ThermoScience}
there was a drive to establish the existence of superfluidity which is
more difficult on the BCS than on the BEC side \cite{Grimm,Jin3} of the
resonance.  There has been a sequence of experiments which have
effectively made this case, beginning first with fast sweep experiments
\cite{Jin4,Ketterle3}, then thermodynamical measurements
\cite{ThermoScience}, and finally detection of quantized vortices
\cite{KetterleV}. We discuss the first two methodologies here in the
context of our theoretical framework.  We limit our discussion first to
the case of unpolarized superfluids.

\begin{figure}[tb]
\centerline{\includegraphics[]{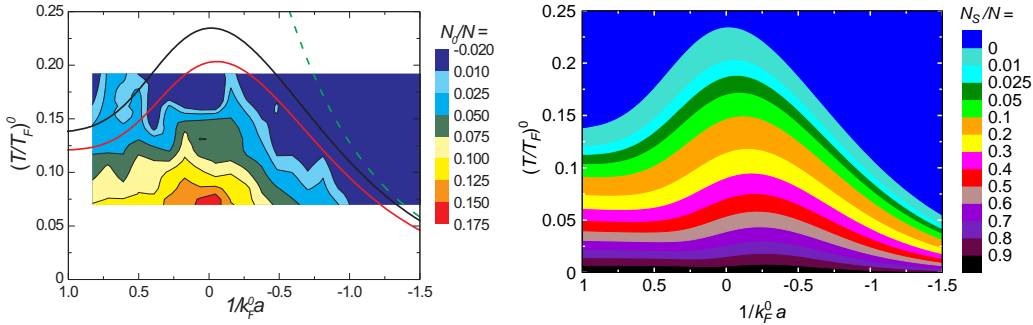}}
\caption{(Color) Earliest evidence for
superfluidity: phase diagram of $^{40}$K as a function of $(T/T_F)^0$
and $1/k_F^0a$. This compares both experiment (left panel) and theoretical
computations (right panel). In the left panel, a contour plot of the
measured condensate fraction $N_0/N$ as a function of $1/k_F^0a$ and
effective temperature $(T/T_F)^0$ is compared with theoretically
calculated contour lines at $N_s/N=0$ (at the superfluid transition,
black curve) and 0.01 (red curve).  The experimental data have an
overall systematic uncertainty of approximately 0.1 in $1/k_F^0a$.
The experimental contour at $N_0/N=0.01$ and the
theoretical line for $N_s/N=0.01$ are in good agreement.  The dashed
line represents the naive BCS result $T_c/T_F^0 \approx 0.615
e^{\pi/2k_F^0a}$. The right panel represents a more complete
theoretically computed equilibrium phase diagram, with contour lines
for $N_s/N$.  Here all temperatures are measured in the Fermi gas
regime. From Ref.~\cite{Jin_us}.}
\label{fig:Phase}
\end{figure}

The left panel in Fig.~\ref{fig:Phase} is a plot of the first phase
diagram representing the condensate fraction \textit{vs} $1/k_F^0a$, as
obtained in Ref.~\cite{Jin4} for $^{40}$K. Subsequently, similar studies
\cite{Ketterle3} were undertaken for $^6$Li. The figure is based on
starting the system off in the free Fermi gas regime where it can be
associated with an initial known temperature $(T/T_F)^0$, and then
adiabatically sweeping closer to unitarity. [Here $T_F=T_F^0$,
$k_F=k_F^0$ and $(T/T_F)^0$ are all measured in the noninteracting Fermi
gas limit]. Once the near-unitary gas (of fixed entropy) is obtained, a
fast sweep is made to the BEC regime, where the condensate fraction can
be read off from a bi-modal profile.  The presumption here, for which
there is considerable experimental support \cite{Jin4,Ketterle3}, is
that, even if the condensate fraction is not conserved upon a fast
sweep to BEC, the presence or absence of a condensate will be preserved.
The time frame for the sweep will not allow a condensate to form in the
BEC if there were none present near unitarity, nor will it allow a
condensate to disappear if it was present initially.

The lines drawn on top of the experimental contour plots are the
calculated \cite{Jin_us} condensate fraction contours as a function of
the adiabatic sweep-projected temperatures $(T/T_F)^0$ for a 0\% and 1\%
condensate fraction. These essentially correspond to the
normal-superfluid phase boundary which is expected to be rather well
measured in these sweep experiments.  The figure to the right presents a
full plot of the theoretical condensation fraction, importantly,
measured with respect to the adiabatic sweep-projected temperatures
$(T/T_F)^0$.  This, then, is the theoretical phase diagram.  For the 1\%
case, the overall trends yield good agreement between theory and
experiment, except for the small ``overshoot" (of unknown origin) which
appears in the BEC side of the data.

\begin{figure}
\centerline{\includegraphics[width=2.8in,clip]{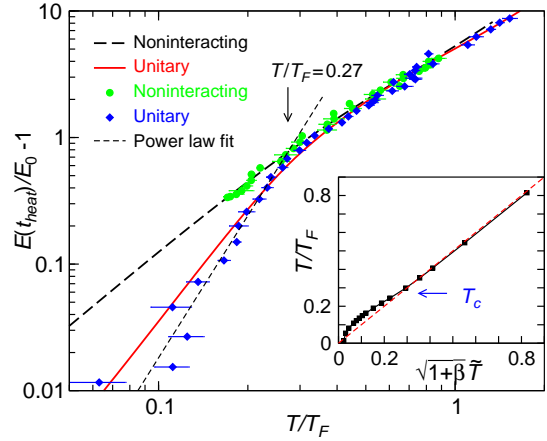}}
\caption{Evidence for a superfluid phase via plots of the energy $E$ vs
  physical temperature $T$. The upper curve (long dashed line) and data
  points correspond to the BCS or essentially free Fermi gas case, and
  the lower curve and data correspond to unitarity. The latter provide
  indications for a phase transition via a slope change. The thin dashed
  line represents a power law fit to the unitary data below the
  transition.  The inset shows how temperature must be recalibrated
  below $T_c$. Here $\beta$ is a parameter which characterizes the
  interaction energy at unitarity for $T=0$, and $\tilde{T}$ is a
  empirical temperature scale.  From Ref.~\cite{ThermoScience}.}
\label{fig:231new}
\end{figure}

The second generation experiments which helped to establish
superfluidity were based on thermodynamical measurements.  In
Fig.~\ref{fig:231new} we show data from Ref.~\cite{ThermoScience}.  What
is plotted is the measured energy as a function of temperature on a
log-log scale.  This temperature represents a theoretical recalibration
of an effective profile-measured temperature $\tilde{T}$. The key
feature here is that the data (indicated by the points) show an abrupt
change at a temperature one can call $T_c$. This abrupt change occurs
for the unitary scattering case. No such feature is seen for the
noninteracting Fermi gas, also plotted in the figure.  Furthermore, this
abrupt change in $E$ vs $T$ is seen \textit{even in the raw data points} (not
shown here), without appealing to a theoretical re-calibration.  All of
this is very suggestive of a specific heat discontinuity, which is to be
associated with a phase change -- presumably to a superfluid phase.

The inset in the figure shows how the effective temperature $\tilde{T}$
which is obtained from a Thomas-Fermi fit to the calculated density
profile compares with the physical temperature $T$.  The inset shows
that a recalibration is necessary only below $T_c$, to account for the
presence of a condensate.  We stress that the observation of a phase
change made by the Duke group \cite{ThermoScience} is not dependent on
this recalibration.  Hence these experiments provide good evidence for a
transition between a normal and superfluid phase.

The last generation experiment to make the case for superfluidity was
the observation of quantized vortices by the MIT group
\cite{KetterleV}. These very convincing experiments will be discussed
elsewhere in this lecture series.

\section{Fermi Gases with Imbalanced Spin Population}

The latest excitement in the field of trapped fermions pertains to gases
with deliberately imbalanced spin populations
\cite{ZSSK06,PLKLH06,ZSSK206}.  In large part this is motivated by
interest from theorists in other disciplines such as dense QCD and
(isospin asymmetric) nuclear matter
\cite{Wilczek,LW03,FGLW05,Sedrakian}.  From the condensed matter
viewpoint there has been an underlying interest in exotic and 
intriguingly
elusive phases associated with Zeeman effects in superconductors, such
as that proposed by Larkin and Ovchinnikov and by Fulde and Ferrell
(LOFF).  In the LOFF state \cite{FFLO} the condensate has a net momentum
of a fixed $\mathbf{q}$ or of $+\mathbf{q}$ and $-\mathbf{q}$.  Even
more elaborate crystalline lattices of various $\mathbf{q}_i$ have also
been contemplated \cite{LOFF_Review}.  An additional, and very important
motivation for these studies is associated with the recent claims
\cite{ZSSK206} that when there is a population imbalance, the density
profiles will indicate whether or not superfluidity is present and they,
moreover, provide an internal mechanism for thermometry.
Thus, because these experiments are claimed to identify $T_c$ itself, and
because they pertain to thermometry, it should be clear that a
theoretical analysis of these experiments requires an understanding of
the effects of $T \neq 0$.

We begin our discussion by summarizing some key experimental
observations \cite{ZSSK06,PLKLH06,ZSSK206}.  In a trapped cloud there
appears to be a superfluid core, which, at the lowest temperatures, is
unpolarized.  Outside of this core there is a normal region, in which
both spin components are present and this carries a significant fraction
of the polarization.  Beyond this ``mixed, normal region" there is a
free Fermi gas composed of the majority spin species, which carries
additional polarization.  At suffiently low temperatures, there appears
to be a form of phase separation in which the superfluid and normal
phases are associated with zero and finite polarization, respectively.

We turn now to a theoretical understanding \cite{ChienRapid,Chien_prl}
of these experiments at $T \neq 0$ first for the case of a homogeneous
gas, and then later in the trapped configuration.  There has been some
work along these same lines elsewhere in the literature \cite{YD05},
although without incorporating noncondensed pair effects. We focus on
the Sarma or breached pair phase \cite{Sarma63}.  The figures we present
do not include the more exotic LOFF or phase separated states. The
formal structure for addressing the former \cite{LOFFlong} is very
similar to that of the Sarma state; we will defer a brief discussion of
these until the end.  To help with the clarity of the presentation, we
state our major conclusions for the homogeneous and trapped
configurations in the unitary regime, at the outset.
\begin{itemize}
\item In the \textit{homogeneous} case temperature serves to stabilize
  the polarized superfluid phase. Such a phase is unstable in the ground
  state.
\item In a \textit{trap} at low $T$, the superfluid core remains
  unpolarized.  Except at very low $T$, spin polarization will be
  mostly carried by fermions coexisting with strongly interacting,
  thermally excited noncondensed pairs which appear outside the
  core. Thus, pseudogap effects, which have been emphasized throughout,
  are very important here.
\item With increasing $T$, polarization tends to continuously penetrate
  into the superfluid core of the trap, until at, or even below $T_c$,
  the polarization is uniformly distributed throughout the cloud.
\item This superfluid Sarma or breached pair phase is limited in the amount of
  polarization it can accommodate, especially near unitarity.  
  This applies to both traps and homogeneous sytems.
\end{itemize}

\begin{figure}
\centerline{\includegraphics[clip,width=3.8in]{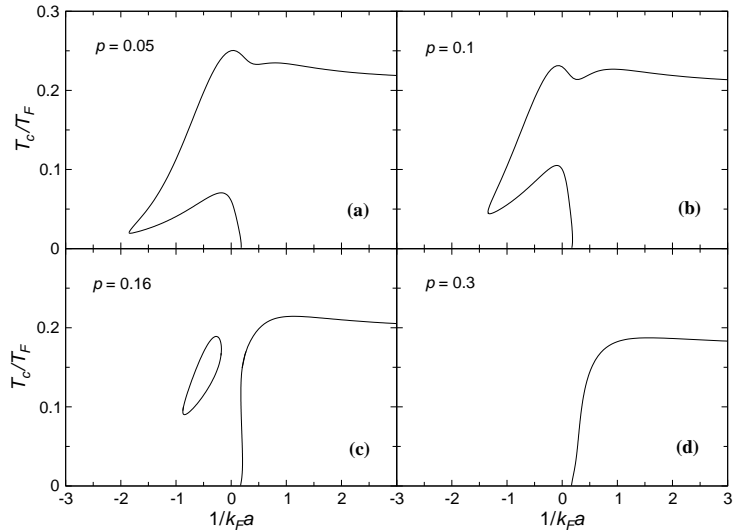}}
\caption{Typical behavior of $T_c$ as a function of $1/k_Fa$ for spin
  polarizations $p=0.05$, 0.1, 0.16, and 0.3. The $T_c$ curve splits
  into two disconnected curves around $p=0.14$. This figure should be
  compared with left panel of Fig.~\ref{fig:Tc} for the unpolarized
  case. From Ref.~\cite{Chien_prl}.}
\label{fig:TcImb}
\end{figure}

Figure \ref{fig:TcImb} presents a plot of $T_c$ for a homogeneous system
as a function of $1/k_Fa$ for various polarizations.  This figure should
be compared with the left panel of Fig.~\ref{fig:Tc}.  With this
comparison, one sees at once that there can be no superfluidity in the
deep BCS regime, once the polarization is different from zero. And when
superfluidity first appears on the BCS side of resonance it is
associated with two $T_c's$ at given $1/k_Fa$. The larger the
polarization $p$, the harder it is for a homogeneous system to support
superfluidity, except in the BEC regime.

\begin{figure}
\centerline{\includegraphics[clip,width=3.in]{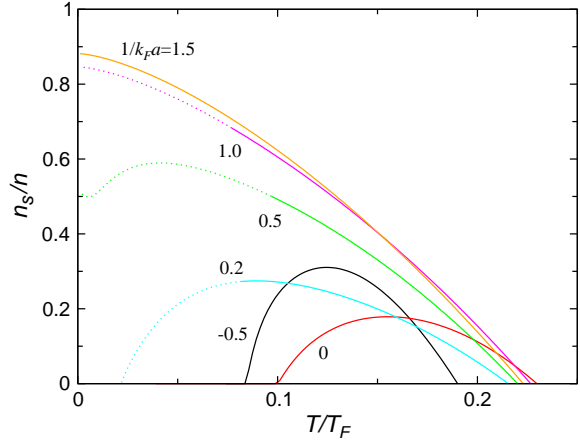}}
\caption{ Normalized superfluid density $n_s/n$ as a function of $T/T_F$
at $p=0.1$ for various $1/k_Fa$ from BCS to BEC.  The dotted (segments
of the) curves represent unstable solutions. The fact that at resonance
and on the BCS-side there are 2 $T_c's$ is consistent with the previous
Figure. From Ref.~\cite{Chien_prl}.}
\label{fig:ns}
\end{figure}

To understand the meaning of these two $T_c$'s we plot the superfluid
density $n_s(T)$ as a function of $T$ in Fig.~\ref{fig:ns}, for several
different values of $1/k_Fa$.  If one focuses on the unitary case, for
definiteness, one can see that $n_s$ vanishes at two different
temperatures. The lower $T_c$ corresponds to the onset of superfluidity.
At temperatures below this, the (breached pair or Sarma) state is
unstable.  Similarly, at the upper $T_c$, superfluidity disappears in
the usual way; it is destroyed by thermal excitations.  For positive
values of $1/k_Fa$, the figure shows that the $n_s(T)$ curves stop
abruptly. This is indicated by the dotted segments of the curves which
represent thermodynamically unstable solutions.

\begin{figure}
\centerline{\includegraphics[clip,width=3.in]{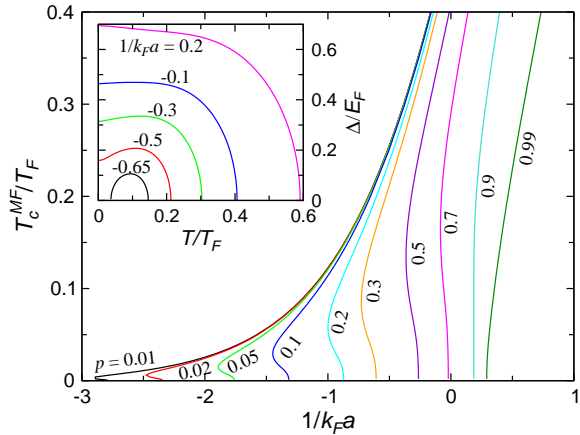}}
\caption{Mean-field behavior of $T_c^{MF}$ as a function of $1/k_Fa$ for
different spin polarizations $p$.  Shown in the inset is the pairing gap
$\Delta(T)$ at different $1/k_Fa$ for $p=0.3$ which can vanish at two
distinct temperatures.  Here $E_F\equiv k_B T_F \equiv\hbar^2 k_F^2/2m$
is the noninteracting Fermi energy in the absence of polarization.
The quantity $T_c^{MF}$ represents the pairing onset temperature
$T^*$.}
\label{fig:TcMF}
\end{figure}

The fact that $T_c$ may be double-valued could have been anticipated in
a simpler set of calculations performed at the strict mean field level,
and discussed elsewhere in the literature \cite{Sedrakian}.  Here one
solves Eqs.(\ref{eq:11}) and (\ref{eq:12}) only, without imposing
(\ref{eq:sum}) and (\ref{eq:delta_pg}).  In Fig.~\ref{fig:TcMF}, we
present a plot of $T_c^{MF}$ as a function of $1/k_Fa$ for a range of
$p$.  This quantity can be viewed as the pairing onset temperature
$T^*$, when there is a stable superfluid phase. In the inset of
Fig.~\ref{fig:TcMF} we plot $\Delta(T)$ at different $1/k_{F}a$ for
$p=0.3$.  For $p<0.9$ and sufficiently low $T_c^{MF}$, we find that
there are two $T_c^{MF}$ values.  This structure implies that $\Delta$
is nonmonotonic \cite{Sedrakian} in $T$, as indicated by the bottom
curve in the inset of Fig.~\ref{fig:TcMF}. The two zeroes of $\Delta$
represent the two values of $T_c^{MF}$.  In contrast to the more
conventional behavior (shown in the top curve for stronger pairing
interaction), $\Delta$ \textit{increases} initially with $T$ at low
temperature when $1/k_{F}a$ is sufficiently small.

We summarize by noting that these results indicate that
\textit{temperature enables pairing associated with the breached pair or
Sarma state}. This was also inferred in Refs.~\cite{YD05} and
\cite{Sedrakian}.  In general superfluids, one would argue that these
two effects compete.  We may view, then, this unusual polarized phase as
an ``intermediate temperature superfluid".

\begin{figure}
\centerline{\includegraphics[width=4.9in,clip]{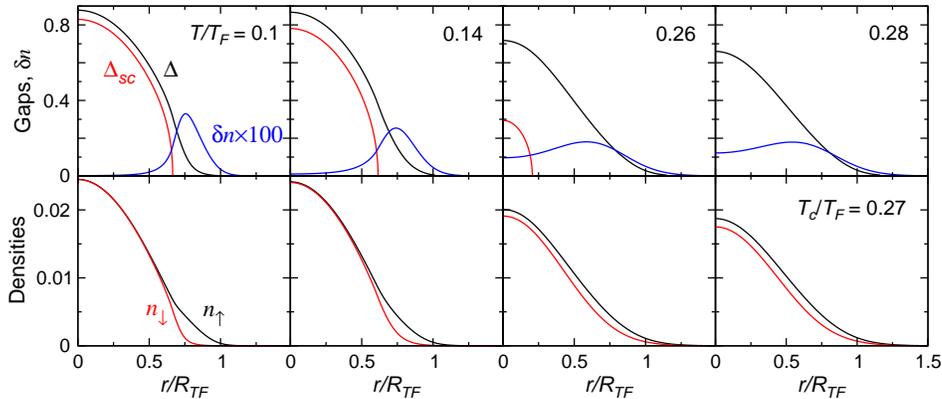}}
\caption{Spatial profiles of the Gap ($\Delta$), order parameter
  ($\Delta_{sc}$), and the density of the up ($n_\uparrow$) and down
  ($n_\downarrow$) spin components and their difference ($\delta n$) for
  a unitary gas in a trap at $T/T_F=0.1$, 0.14, 0.26, and 0.28, from
  left to right. Except at very low $T$, most of the polarization is
  carried by fermions co-existing with noncondensed pairs,
  \textit{i.e.}, in the pseudogap regime where $\Delta_{sc} = 0$, but
  $\Delta \neq 0$. Here $T_c/T_F\approx 0.27$.}
\label{fig:15}
\end{figure}

We now turn to the behavior of these superfluids in a trap.  The same
calculations are applied to the Sarma or breached pair state using the
LDA to incorporate trap effects. Figure \ref{fig:15} shows the resulting
behavior at unitarity for polarization of 15\% and for various
temperatures from below to just above $T_c$.  The upper panels plot the
order parameter $\Delta_{sc}$ and the (total) gap parameter $\Delta$.
Superposed on these plots is the polarization $\delta n$.  The lower
panels present the density profiles for each spin state.  Several
important features can be gleaned from the upper panels. At the lowest
temperatures the bulk of the polarization is in a region where
$\Delta_{sc} =0$, but $\Delta \neq0$; thus polarization is excluded at
low $T$ from the superfluid core.  Moreover, it can be seen that an
excitation gap $\Delta$ is present throughout most of the cloud.
Whenever $\Delta \neq 0$ one can infer that both spin states are
present.  For non-zero $\Delta$, the particle profiles are necessarily
different from those of a non-interacting gas.  The bulk of the
polarization appears in the ``normal, mixed region", and within this
portion of the trap there are strong interactions between the two spin
states.  Only at the very edge of the trap is there an exclusively
majority component (and here $\Delta =0$).  This region is occupied by a
non-interacting Fermi gas and can thereby be used to set the temperature
scale for the trapped cloud.

\begin{figure}
\centerline{\includegraphics[width=4.4in,clip]{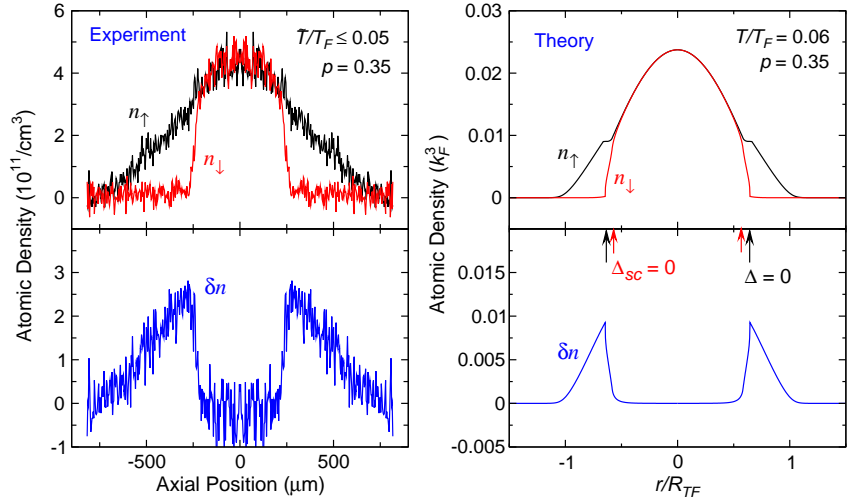}}
\caption{Comparison of theoretically computed density distributions of
  the up and down spin components (upper row) and their difference
  (lower row) at unitarity with the Rice group data. The central issue
  here is not the quantitative comparison, but rather that (as seen in the
  theoretical curves) polarization is confined to outside the superfluid
  core. In the Rice data, $n_\downarrow$ has a narrower distribution due
  to non-LDA effects. Data from Ref.~\cite{Rice_new}.}
\label{fig:8}
\end{figure}

\begin{figure}
\centerline{\includegraphics[]{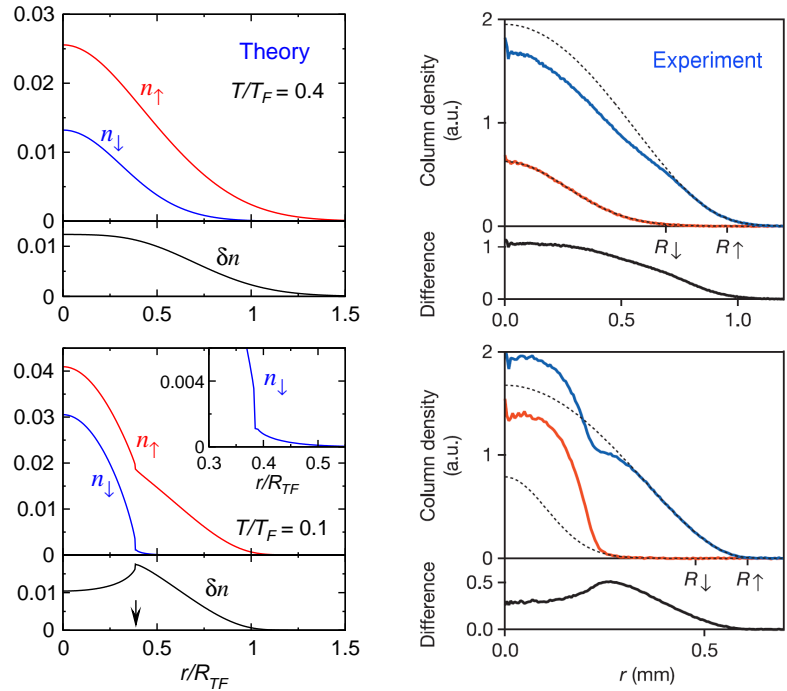}}
\caption{Comparison of theoretically computed density distributions of
  the up and down spin components and their difference $\delta n$ at
  different temperatures in the BEC regime with MIT data from
  Ref.~\cite{ZSSK206}, for $p=0.6$. Upper/lower panels are above/below
  $T_c$.  The arrow in the bottom theoretical curve for $\delta n$
  indicates the condensate edge which is reflected in features in both
  $n_{\uparrow}$ and $n_{\downarrow}$, much as in the data.}
\label{fig:1}
\end{figure}

We end this Section with some comparisons with experiment.  We address
qualitative effects in Fig.~\ref{fig:8} and semi-quantitative effects in
Fig.~\ref{fig:1}. The former is for the unitary case and the latter is
in the BEC regime.
The left panels in Fig.~\ref{fig:8} show data from the Rice group
Ref.~\cite{Rice_new}. The upper figure plots the density profiles for
each spin state and the lower panel, their difference.  The unpolarized
core is evident, as is the sharp edge beyond which polarization abruptly
appears.  This behavior has been interpreted \cite{PLKLH06,Rice_new} as
suggesting phase separation.

Because the aspect ratios for the trap are not maintained in the
profiles, it has been argued \cite{PLKLH06,Rice_new} that the LDA scheme
may not be appropriate for addressing experiments on these highly
anisotropic traps. In the experimental data, non-LDA effects lead to a
narrower distribution for $n_\downarrow$ along the axial direction while
it is broadened in the radial direction.  Despite this caveat we plot
our counterpart theoretical results at $T=0.06T_F$ (right panels) for
qualitative comparison.  This plot is designed primarily to introduce
theoretical observations (which can be superposed, in effect, on the
experimental plot) concerning where the superfluid phase resides in the
trap.  This knowledge cannot be directly gleaned from these particular
experiments. These low $T$ theoretical results are rather striking, for
they make it clear that when the order parameter is present
 the polarization is excluded.  Rather, the
polarization appears in the Fermi gas region outside the condensate
core.  At higher $T$ and lower $p$ (e.g, $T=0.1T_F$, $p=0.15$), the
polarization is carried largely within the ``pseudogap" regime, where
there are strong pairing correlations, but no long range order.

We turn finally to a comparison of data from the MIT group
\cite{ZSSK206}, and in the BEC regime.
The experimental data (for polarizations near 0.6) are plotted on the
two right panels which show the particle density profiles for each spin
state and their difference $\delta n$. The upper panels correspond to
temperatures which are believed to be in the normal state, whereas, the
lower panels are for the superfluid phase. The lightly dotted curves in
the experiment represent extrapolations of the Thomas-Fermi fits to the
curves at large radii.  In the superfluid phase, the data show that the
polarization at the core center is considerably smaller than it is for
the normal state.  One apparent difference between above and below $T_c$
experiments, is that at low $T$ (i) the minority component has
contracted into the center of the trap. (ii) Another signature is a kink
in the majority profile. Yet another signature is that (iii) there is a
clear bi-modality in the minority component.

This behavior can be compared with the theoretical results plotted in
the left two panels above and below $T_c$ for a polarization $ p=0.6$.
The kink in the majority component at low $T$ can be clearly attributed
to the edge of the condensate as denoted by the vertical arrow (where
$\Delta_{sc}=0$).  The bi-modality in the minority component is
amplified in the lower inset to make it more evident.  These theoretical
plots, at a qualitative level, exhibit the three central features of the
data noted above.

\section{Conclusions}

Throughout this Review we have stressed that temperature is important in
understanding and characterizing the ultracold Fermi gases.
Experimentally, one is never strictly at $T=0$; as a result there are
thermal excitations of the gas which need to be characterized both
theoretically and experimentally.  At a formal level, we have discussed
how temperature enters into the standard BCS-BEC crossover theory by
leading to a distinction between the superconducting order parameter
$\Delta_{sc}$ and the gap parameter $\Delta$.  This difference reflects
the existence of pre-formed pairs above $T_c$ and noncondensed pairs
below.

At a more physical level we have shown how temperature changes the
character of the gas from a superfluid at low $T $ to an unusual (but
strongly interacting) normal fluid at moderate $T $, and to ultimately a
free Fermi gas at high $T$.  This strongly interacting normal fluid is
most interesting, for unlike the Landau Fermi liquid (or gas) which
exists just above $T_c$ in a strictly BCS superfluid, here there is a
normal state excitation gap or pseudogap.  By looking at three different
experiments, we have provided evidence for this normal state excitation
gap (in RF pairing gap spectroscopy \cite{Grimm4} and in thermodynamics
\cite{ThermoScience} and for the below-$T_c$ counterpart (noncondensed
pairs), via the shape of the particle density profiles
\cite{Thomas,Grimm2}.

In this Review we have discussed how temperature can be measured through
adiabatic sweep thermometry \cite{Jin4} and through Thomas-Fermi fits to
density profiles \cite{Thomas,ThermoScience}.  We have shown how
temperature played an important role in establishing the first
generation and earliest evidence for superfluidity -- based on fast sweep
experiments which yield \cite{Jin4} a phase diagram in the
temperature vs $1/k_Fa$ plane.  It also played an important role in the
second generation indications for superfluidity associated with
thermodynamical measurements \cite{ThermoScience}.

Finally, we have addressed a new and exciting class of experiments
\cite{ZSSK06,PLKLH06,ZSSK206} involving spin population imbalance and
shown that here too temperature plays a critical role.  From an
experimental point of view, the most exciting features of these
experiments are that they show how the density profiles can be used to
establish the transition temperature $T_c$. Moreover, through the wings
of the profiles, they provide a theory-independent mechanism for
thermometry.  From our theoretical perspective, which focuses on
temperature \cite{Chien_prl,ChienRapid}, what is also exciting is that
(i) the stability and character of these polarized superfluids is very
sensitive to temperature and (ii) even in the normal state one sees
strong interactions between the two spin components, which we associate
with the finite $T$ pseudogap effects we have been discussing in this
Review. As stated in Ref.~\cite{ZSSK206}, ``Already at high
temperatures, above the phase transition, the larger [majority] cloud's
profile is strongly deformed in the presence of the smaller [minority]
cloud, a direct signature of interaction."

Our discussion, thus far, has not addressed phase separation \cite{SR06}
or the exotic LOFF states \cite{FFLO,Combescot} associated with
polarized superfluids. The latter, which correspond to a condensate with
a finite momentum $\mathbf{q}$, may well be ground states of these spin
polarized gases for some range of scattering lengths $a$, since the
polarized BCS like phase (with $\mathbf{q}=0$) is only stable at
intermediate temperatures.  These LOFF phases are discussed elsewhere in
this volume. There are arguments in the literature that in some more
general form they play an important role near unitarity
\cite{Machida2,Kinnunen} in trapped gases.  At a theoretical level, LOFF
states may be included \cite{LOFFlong} following the formalism we have
outlined in this Review.

Underlying the interest in this general field of ultracold gases is the
possibility that they may shed light on the high temperature cuprate
superconductors either in the context \cite{ourreview,LeggettNature} of the 
BCS-BEC crossover scenario
or in the context of optical lattices and Hubbard model simulations.
For the former, one can return to the question of where they would fit
on a phase diagram plot (such as that presented in Fig.~\ref{fig:Tc}
which addresses $T_c$ vs $1/k_Fa$).  When the calculations are properly
redone for $d$-wave pairing on nearly two dimensional lattices, the
values and shape of the $T_c$ curve are in quite good agreement with
experiment \cite{Chen2,ourreview,ReviewJLTP}. Moreover, $T_c$ vanishes
well before the BEC limit is reached \cite{Chen1}; one can, then,
deduce that in this scenario
the cuprates are close to the unitary regime, just as are the
ultracold gases.

Whether or not this crossover picture turns out to be correct for high
$T_c$, another very important rationale for its study is the possibility
of generalizing what is undoubtedly the most successful theory in
condensed matter physics: BCS theory.  We now understand from the
ultracold gases that the nature-made superconductors to which the
original theory has been applied, are only a very special case of a much
more general class of superfluids.  And it is most fitting on this
fiftieth anniversary of BCS that we pay homage to this most remarkable
of theories by recognizing its even greater generality.

\acknowledgments 

We wish to warmly thank all our past collaborators who have contributed
to the work and figures presented here: Jelena Stajic, Yan He and
Chih-Chun Chien, as well as John Thomas, Joe Kinast and Andrey Turlapov,
as well as Murray Holland, Marilu Chiofalo and Josh Milstein, as well as
Debbie Jin, Cindy Regal and Markus Greiner. This work was supported by
NSF PHY-0555325 and NSF-MRSEC Grant No.~DMR-0213745.


\end{document}